# Sp.A.C.E. Lab.

Verma, D., Mumm, O. and Carlow, V.M., 2023. "Generative agents in the streets: Exploring the use of Large Language Models (LLMs) in collecting urban perceptions" arXiv preprint.

## Journal Paper Preprint

# Generative agents in the streets: Exploring the use of Large Language Models (LLMs) in collecting urban perceptions

Deepank Verma[1,*], Olaf Mumm[1], Vanessa Miriam Carlow[1]

[1] Institute for Sustainable Urbanism—Spatial Analytics & Cross-Disciplinary Experimentation Lab (ISU SpACE Lab), Technische Universität Braunschweig, 38106 Braunschweig, Germany

* Corresponding author: d.deepank@tu-braunschweig.de

**Abstract:** Evaluating the surroundings to gain understanding, frame perspectives, and anticipate behavioral reactions is an inherent human trait. However, these continuous encounters are diverse and complex, posing challenges to their study and experimentation. Researchers have been able to isolate environmental features and study their effect on human perception and behavior. However, the research attempts to replicate and study human behaviors with proxies, such as by integrating virtual mediums and interviews, have been inconsistent. Large language models (LLMs) have recently been unveiled as capable of contextual understanding and semantic reasoning. These models have been trained on large amounts of text and have evolved to mimic believable human behavior. This study explores the current advancements in Generative agents powered by LLMs with the help of perceptual experiments. The experiment employs Generative agents to interact with the urban environments using street view images to plan their journey toward specific goals. The agents are given virtual personalities, which make them distinguishable. They are also provided a memory database to store their thoughts and essential visual information and retrieve it when needed to plan their movement. Since LLMs do not possess embodiment, nor have access to the visual realm, and lack a sense of motion or direction, we designed movement and visual modules that help agents gain an overall understanding of surroundings. The agents are further employed to rate the surroundings they encounter based on their perceived sense of safety and liveliness. As these agents store details in their memory, we query the findings to get details regarding their thought processes. Overall, this study experiments with current AI developments and their potential in simulated human behavior in urban environments.

## 1. Introduction

### *1.1. Related work in urban perception*

The way humans perceive their environment and act is deeply ingrained and personal. Individuals' behavior results from their immediate surroundings and the pursued activity with a particular target in mind (Haggard, 2017; I. Altman, 1976; Ittelson, 1978). The environment influences and stimulates notions and thoughts that define their experiences of the surroundings (Berlyne, 1951; Montserrat Degen & Rose, 2012). The journeys one undertakes to fulfill objectives depend on many decisions. Due to their complexity, these decisions are often difficult to quantify, as cognitive thoughts and perceptions are challenging to evaluate and reproduce in experiments.

Researchers have tried to isolate environmental characteristics and study behavioral patterns and cognitive inputs from visual and aural stimuli (Carlow et al., 2022; Hillnhütter, 2022; Liu et al., 2013; Verma et al., 2020) and mobility-related decision-making such as wayfinding and route selection (Tong & Bode, 2022; Zomer et al., 2019). Traditionally, as a replacement for in-situ experimentation involving participants in the field, the experiments were carried out with the help of photographs and videos reproduced to represent real-world scenes (Herzog, 1989; Herzog et al., 1976; Kaplan et al., 1972; Nasar, 1987). The participants then judged these scenes in an ex-situ setting based on the environmental attributes being studied. This provided more flexibility in conducting experiments where more subjects could participate in these studies. As the technology progressed, non-interactive photos and videos gave way to virtual reality (VR) mediums (Kuliga et al., 2015; Portman et al., 2015). With VR, researchers can effectively create experimental scenes that block or restrict some environmental features to effectively study the relationship between the desired features and participants' intended cognitive attitudes (R. K. Dubey et al., 2023; Jeon & Jo, 2020; Natapov & Fisher-Gewirtzman, 2016; Shushan et al., 2016). The results from these studies gave the research community believable proxies for real-world relationships between environmental features and human perceptions.

While rich literature originating from these studies benefited a general understanding of human psychology and behavior in urban settings, the implementation suffered due to the complex nature of experiments and less scalable results in large urban contexts (Evans et al., 1982; Harvey & Aultman-Hall, 2016). As an active research domain, efforts are being made to develop scalable research approaches that answer focused questions related to particular behaviors and decisions taken to pursue specific tasks and the role of environmental characteristics in it. However, a consensus on common guidelines and standard practices for undertaking these studies has not been reached thus far.

The recent wave of AI algorithms, mainly in the assessment of visual and auditory scenes and access to street view imageries, revived these studies and pioneered the way to scale the methodology to multiple cities and regions (A. Dubey et al., 2016; Salesses et al., 2013; Santani et al., 2018). These studies successfully linked basic human emotions such as safety, liveliness, and boredom with overall visual and auditory features. The image databases have helped gather links between mental and physical health and the presence of various street features (Kruse et al., 2021; R. Wang et al., 2019). Socioeconomic factors like prosperity, poverty, and crime have also been identified with the multi-temporal usage of these databases (Mooney et al., 2017; Naik et al., 2017).

The quantification of environmental characteristics is made possible by these algorithms; however, the attempts to find ways to gather meaningful responses at scale have been far fewer. Since these studies only included pairwise comparisons and Likert scale ratings to allow large groups of participants and to rate large group images or pairs in less time, not much is studied on the human side of the equation. Also, since the process is anonymous, the demographic details of the participants who rated these photo pairs are difficult to ascertain. While it's a fruitful attempt to gather large-scale human-based information, little information can be extracted from the ideation and thought process involved in providing these assessments.

While the research communities' ongoing search for harnessing the full potential of these AI models has continued, a recent advancement, Large Language Models (LLMs), has been transforming the AI landscape (Zhao et al., 2023). LLMs are advanced language models capable of generating human-like thoughts by observations and planning future steps virtually (Trott et al., 2023). These models' capabilities have allowed them to be utilized across multiple domains, outperforming most previous state-of-the-art AI models.

### 1.2. *Large Language Models (LLMs)*

LLMs, or Large Language Models, are based on Transformer-based DL models capable of summarizing and predicting new content based on the provided context. They replace Natural Language Processing models for sentiment prediction, language translation, etc. (B. Min et al., 2024). With the latest Transformers architecture, models can utilize billions of texts for training. A trained LLM can semantically organize the learned text to provide answers akin to or more efficiently than humans and, more importantly, perform new procedures not hardcoded in the script (Andreas, 2022). These models can be used for conversation as a chat agent, language translation, and general queries. However, since these models are generalized over large text datasets with various topics, generating output for a focused set of activities is complex.

"Prompts" are crafted texts intended to steer the LLM's reasoning in a desired direction (Jin Chen et al., 2023; Shanahan, 2022). For instance, the text "*You are an eminent mathematician, you are given a task to do calculations given the variety of numbers given,*" followed by the numbers, will essentially tell the model to focus on mathematics rather than the entire spectrum of information which it is capable of answering. A prompt has specific parts such as context, instructions, tasks, and recursion (if required). The context sets up the overall intent, which is expected of the model, such as Q&A, role-playing, and expertise. The instructions are a dedicated format in which the model should provide an answer, such as *"in 50 words"*, *"as a table,"* or *"as a 10-year-old"*. Finally, the task is the actual question being asked to the model.

Although Prompt engineering allows the LLMs to generate relevant text according to the user's demand, it still suffers from the burden of its knowledge and tries to provide an answer even when it is slightly made up, often termed a hallucination (Ye et al., 2023). Researchers have recently suggested using the chain-of-thought method, ReAct (Yao et al., 2022), in which the model uses reasoning with itself to provide sensible outputs. This method allows the model to undergo a step-by-step process to question and align itself to provide more accurate results. The ReAct model has been a massive success in Human-Computer Interaction tasks (HCI) involving robots, which are given natural language-based instruction to move and grab objects (Driess et al., 2023).

Owing to their large databases, these models are comparatively huge for consumer-grade computational platforms. Due to this, they are hosted on servers, which can be accessed with the APIs. Although all the proprietary and open-sourced LLMs (Zhao et al., 2023) models follow a similar architecture (Vaswani et al., 2017) for model training, the models differ in the data engineering, types, and amount of datasets on which they are trained. The models balance increased accuracy and high computing resources, defining their applicability in real-world applications.

### 1.3. Equating LLMs and human behavior

While LLMs are not conscious beings and do not possess human-like cognition, they share human characteristics such as contextual understanding, language comprehension, and reasoning and include an extensive knowledge base. Although they do not replace humans, they demonstrate enormous knowledge banks' power and ability to recursively think and adjust their responses based on context (Mahowald et al., 2023). Recent studies showed the capabilities of LLM-based agents as they can adopt a character when given a personality with preferences, a background, and a set of memories (Park et al., 2023; Shanahan et al., 2023). These LLM-powered agents can then be assigned a context emulating the real world and can interact with other such agents to communicate, gather information, disseminate to others, and plan activities. Since these agents are grounded in their little profile and background, they tend to emulate responses, generate new ideas, and indulge in activities. However, since these agents can operate over a large temporal window (Park et al., 2023; G. Wang et al., 2023), the method to store and retrieve essential memories has to be designed by researchers to emulate how humans remember things. Although the agent can emulate human-like behavior and pseudo-feelings, it still cannot understand the vision language and embodiment (Mahowald et al., 2023; Shanahan, 2022). The LLMs are trained with text-based datasets, where even the visual context, such as images and video datasets, are fed as natural language captions. Further, since LLMs do not inherit a human-like sense of locomotion and embodiment from the texts, these models do not understand the directions and movement.

*1.4. Objectives*

This study utilizes the LLMs to simulate human behavior while indulging in two subsequent tasks (Figure 1). The primary task is to study Generative agents' effectiveness in assimilating visual information and a sense of movement in their thought processing to navigate simulated representations of urban areas. We design a Generative agent framework and its visual, movement, and memory modules to gather meaningful observations from the provided visual stimulus. Since the agent can store and retrieve memories, we also converse with the agent to share its decision-making methodology. We create 10 virtual agents, give them different personalities, and allow them to indulge in the task of finding the 'restaurant' in the given area. The groundwork in creating visual and memory modules helps in conducting another task aimed at rating the common perceptual attributes, safety, and liveliness of the visual scenes by these agents. The study uses Google Street View (GSV) imagery and fundamental vision models to obtain visual details, while ChatGPT-3.5 (OpenAI, 2023) is utilized as LLM. The model framework and associated details are also published in the code repository (*https://github.com/deepankverma/gen_agent_langchain.git*).

## 2. Generative agents

The Generative agents are the virtual agents capable of interacting with the virtual environmental details, processing the information, and reacting accordingly. Since LLMs form their brains, the capabilities of such agents are numerous. While the term 'agent' may sound similar to those used in Agent-Based Modeling (ABM), they are fundamentally distinct. ABM has been utilized in simulating human systems, urban growth models, and decision-making research (Hager et al., 2015), providing forecasts, projections, and other non-linear outputs (Bonabeau, 2002). The agents in these models are hard-coded, where the activities and steps taken by the agents are introduced as a condition-based function. Since these agents cannot react differently to what is already coded, the outputs are expected to result from their local interactions, which give results to global patterns. On the contrary, Generative agents can generate reactions and behavior depending upon the given context and prompt management without being explicitly programmed (Andreas, 2022; L. Wang et al., 2023). While agent-based modeling has been a time-tested and scientifically proven method, Generative agents have only recently been proposed, and they are yet to be tested in other scientific domains for validity or usefulness.

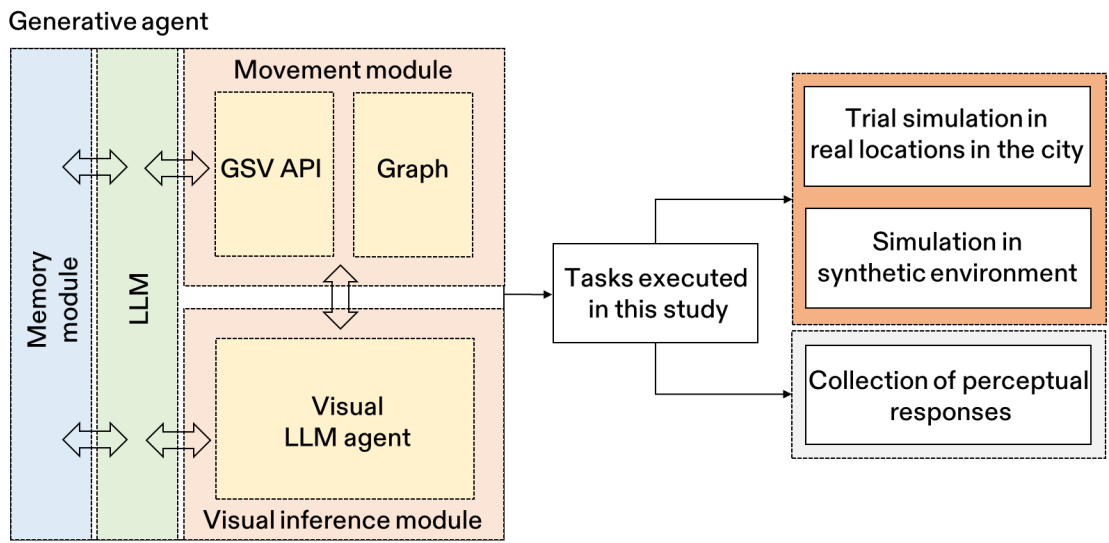

*Figure 1 An overview of Generative agent and performed tasks in this study.*

Although Generative agents are not hard-coded, they need to be provided a context in which they operate, the environment they interact with, the reactions, and the findings' storage and retrieval. In short, the helper functions need to be defined for the smooth functioning of the agents and methods to assess any scientific results. This study utilizes core concepts from the latest research (Park et al., 2023) in this domain, which focuses on gathering insights into human behavior through agent interactions and builds experiments on top of the Generative agent architecture.

### 2.1. Agent architecture and memory management

The agent functions with the help of a defined framework (Figure 1), which includes (1) a memory module that helps the agent to perceive the provided environment and store the relevant details, which can be retrieved later by the agent if necessary and allows it to act, reflect and plan based on the memories it collects. (2) The LLM serves as a backbone of the agent, which acts as a vast knowledge base, helps the agent make decisions, and further converts the communication between other modules into sensible text. (3) The movement module hosts a graph that emulates the real intersections and routes that procure imagery data from GSV API. (4) The visual inference module hosts fundamental visual models that provide details when the image data is given. In this study, the agent architecture and the following modules are designed on top of the *langchain* library in Python, providing a barebones structure for creating Generative agents. The knowledge base for the architecture and memory is borrowed from (Park et al., 2023).

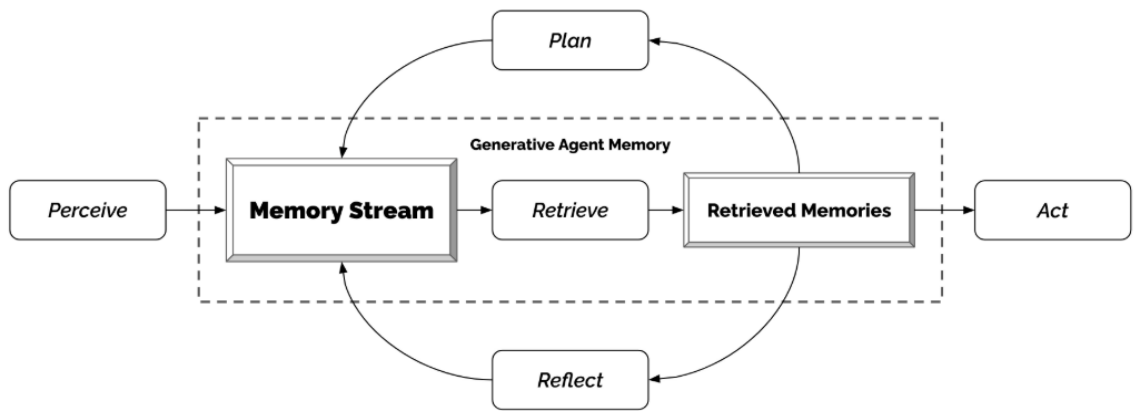

*Figure 2 Image showing memory management flowchart (Park et al., 2023).*

One of the limitations of the LLMs is the number of tokens; simply put, they are the amount of characters that can be communicated with the LLM server at a time. The number of tokens depends upon the LLM architecture being used. Combining the context prompt with the explanation to the agent regarding the task it is supposed to undertake and the following agent observations account for a large set of characters for LLMs to process. Therefore, these observations have to be offloaded from the context window and stored periodically in a memory stream, which can be retrieved by the agent while interviewing it regarding decisions made along the way.

The agents' memory can be created with observations, reflections, and recall or retrieval (Park et al., 2023), akin to humans (Figure 2). The observations are information from the perceived real-world scenes stored in a memory stream. The retrieval of memories from the storage is done as a function of "salience," "recency," and "importance." The agent determines the importance when it rates the information from 1 (most common urban scene) to 10 (most memorable scene) before saving it. The memory module uses recency to tag the time the agent observes the information. This study does not foray into the longitudinal assessment or inter-agent conversations and routine activities; hence, the agent utilizes only the "importance" factor to retrieve the memory and stores it according to the recency. The working of the memory module and the "importance" factor can be seen in Suppl 1.

## 3. Movement simulation

### 3.1. Movement module using street view imagery

Since it is based on a language model trained with texts, the agent does not consist of the ideation of movement. Without the embodiment, the agent has to be passed a context with which it can frame its believable actions in the real world. Studies have utilized the agents in various contexts, such as robotic arms (Driess et al., 2023), indoor navigation (Huang et al., 2023), and outdoor landmark detection (Shah et al., 2022). Along similar lines, this study uses the movement module, which introduces the bidirectional graph and GSV images. The graph nodes represent the decision points, and the edges include information regarding the directions available for the agent to consider, such as left, right, forward, and backward. The decision points are where the agent must decide the direction to move; they can be compared with the conventional street junction. The agent evaluates its decision by gathering image details from all available directions to which it is free to move. The decision nodes store the street view imagery information in the form of URLs and can be fetched with the help of GSV API.

This virtual movement model is an abstract version of actual human movement, which can be better understood as humans closing their eyes while moving in the street but opening them when they reach a junction. Then, they gather the visual details of all the possible routes they can go while standing at the intersection, select one based on the given target, and then close their eyes until they reach another decision point/node.

Figure 3 (a) shows a small test environment in Amsterdam. The decision points are randomly selected with roughly similar spacing and converted to a bidirectional graph (Figure 3 (b)). The number of decision points can be reduced and increased without affecting the working of the overall simulation. The current size of 33 nodes from A to 1H is taken to minimize our computational challenges and LLM's API budgetary restraints. The information regarding the 'directions' attached to the edges is currently a manual exercise. This is due to the constraints in geoprocessing tools, where automatically assigning directions, such as left, right, forward, and backward, correctly to the edges is difficult. Although the direction information, such as bearing and the heading, can be easily calculated as degrees north and south and attached to the edges in place of directions, these values are difficult to understand by the agent emulating a human-like character.

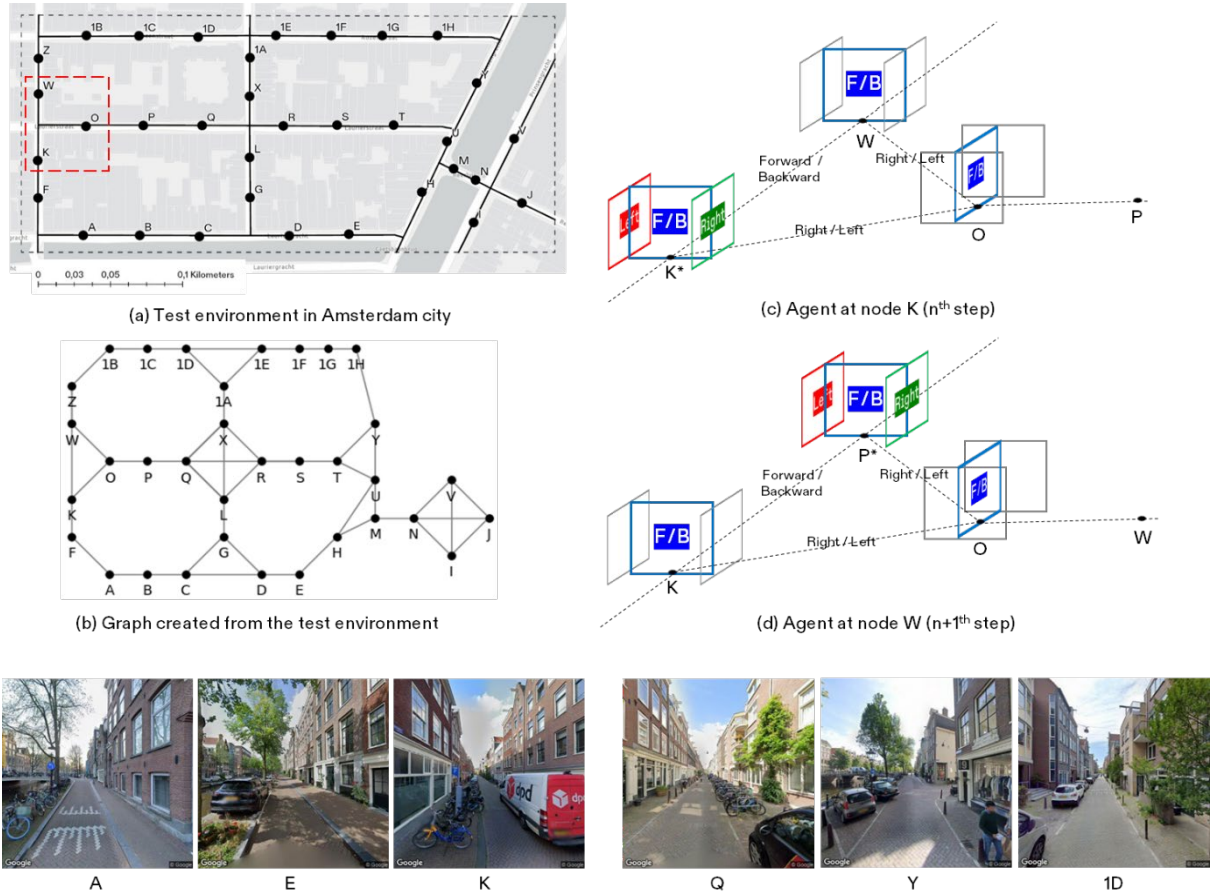

*Figure 3 Bidirectional graph generated from selected locations in Amsterdam.*

The entire process of movement is divided into two steps: (1) When the agent decides which direction to move, associated imagery URLs are recursively supplied to the agent, which downloads front-facing street view images (400px x 400px dimension) of all the available nodes the agent can move to from GSV API. The agent then utilizes a 'visual inference module' to gather details regarding the scenes in these nodes before committing to its preferred path. (2) The GSV API downloads additional left and right-facing images once the agent reaches its decided node (Figure 3 (c) (d)). In this step, the agent tries to find the target assigned, such as a cafeteria, shop, or restaurant, by downloading additional pairs of left and right side images from the API and again utilizing the 'visual module.' These steps are repeated until the agent finds the target assigned. The lat-long coordinates information and the intended heading to feed in the GSV API are generated from OpenStreetMap vector data with the help of geoprocessing tools.

### 3.2. *Environment details through visual inference module.*

The visual inference module deals with extracting high-level details from the images with the help of a group of visual transformer models and Deep Learning (DL) models. The module consists of fundamental visual models such as (1) SegFormer (Xie et al., 2021) semantic segmentation model pretrained on the Cityscapes (Cordts et al., 2016) dataset provides percentage-wise content of streetscape features present in the image, such as trees, buildings, streets, people, vehicles, and sky. The percentage values are converted to high, average, and low for better understanding. (2) Image classification model Wideresnet (Zagoruyko & Komodakis, 2016) finetuned on Places-365 (Zhou et al., 2018) dataset classifies the scene into scene categories commonly found in urban streets and indoors such as restaurants, cafeterias, parks, markets, etc. The model provides the top five classes paired with their confidence. The categories are utilized in the model to frame 'targets.' The model also includes additional information from the top 10 of 102 attributes (Patterson et al., 2014) regarding the nature of the environment, such as no horizon, lighting, enclosed area, man-made, natural, etc. (3) Object detection based on detr-resnet (Carion et al., 2020) model provides a count of detected objects in the scene, mainly number of people, vehicle, lampposts, etc.

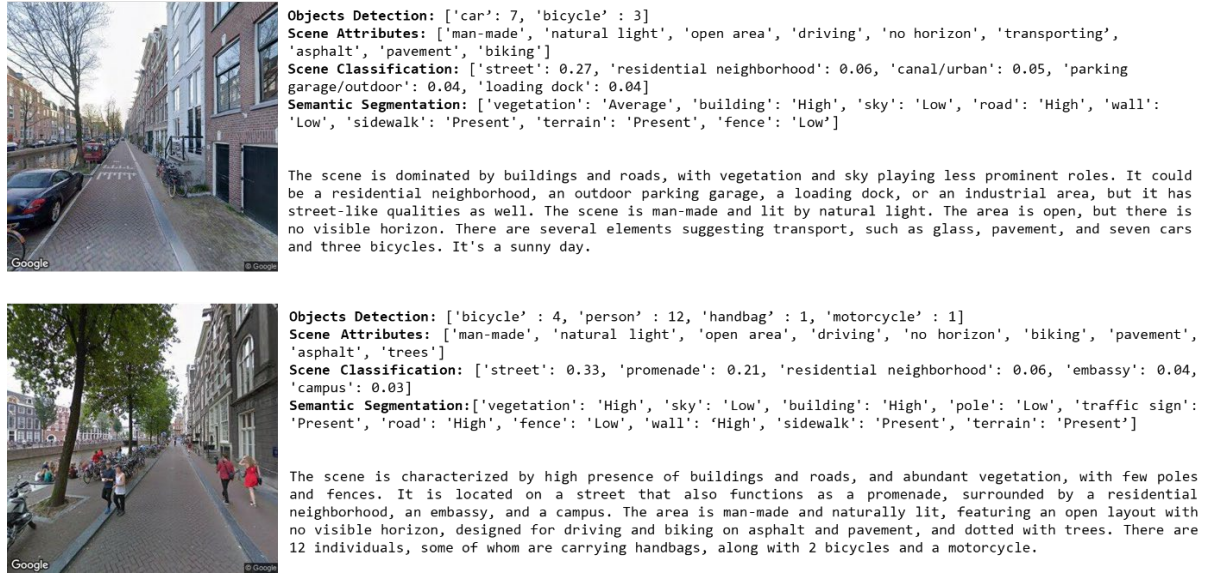

*Figure 4 Details gathered from the imagery and the translation of the data by the LLM agent.*

The visual module is further subdivided into two-step processes where the initial process feeds the street view imagery into visual models to obtain detailed information regarding the scene. The subsequent process utilizes a separate LLM chat agent to convert visual details and values to meaningful summaries appropriate for a general understanding of the technical details from the previous step. Examples of such output are given in Figure 4. The chat agent is tasked with converting all the features and the corresponding values, if any, to meaningful sentences that describe the scene almost in its entirety. This approach is also analogous to recently proposed visual-centric LLM models such as MiniGPT (Jun Chen et al., 2023), Visual ChatGPT (Wu et al., 2023), and GPT-4V (OpenAI, 2023), which can generate image summary utilizing LLMs as a backbone. These models also allow asking specific questions regarding the content of the images. As an alternative to our image understanding and summarization approach, these models can be utilized to achieve similar results. However, these models have been shown to provide oversimplified results and non-existing details (Wen et al., 2023), and their usability in providing precise details in complex images is still being tested by the research community.

### 3.3. Framework testing

We test the designed framework by giving the agent "Max" a starting node 'A' and asking it to find a non-existent 'tree-house' in the area in not more than 15 steps from the starting point. No backstory and traits are added in this trial, and the entire focus is given to the workability of visual and movement modules. The stepwise activity of the agent includes: (1) The agent starts at node 'A,' where the visual inference module processes the street view image linked with the node to provide a list of descriptions, which is further utilized by the LLM visual agent to generate a visual summary of the scene. (2) The agent determines the scene's importance by rating it and storing it in the memory module. The model framework provides new directions/nodes available to the agent to move to. (3) In the next step, the agent moves to the decided node, downloads additional images, and again utilizes a visual module to see if the target "tree-house" is present. (4) Since the target is not present in the area, the agent recursively determines its next move, including the above steps. For each decision, the agent is asked to provide the logic behind the current direction and a reason for not selecting other directions. This gives us an understanding of the thought process behind much of the agent's decision-making. We ran the simulation for 10 runs; the paths traversed by the trial agent in each run are given in Figure 5. Detailed stepwise output is provided in Suppl 1.

Each run begins afresh, which ensures the agent has no memory of previous runs. The paths the agent chooses in each successive run are relatively distinct, showing the in-deterministic nature of the created agent framework. In the current experiment, the agent is not given any personality traits or a backstory, which might help it ground itself in "reality" and make the movement decisions more deterministically in line with the characteristics. Each run is followed by interview questions answered by the agent. The questions included the agent's knowledge of the path and the scenes encountered (Suppl 2). Upon investigating, it is found that the agent can usually track the previous steps and the choices made along the way. Also, it can describe the motive behind choosing particular directions in detail. However, it struggles to fetch the key aspects of the scenes if they have too many details. The agent also fills out the information if it cannot remember. Further, since the agent personality is not being set, the reasons for choosing and not choosing a particular path at a fixed decision point keep changing with every run. Altogether, this exercise provides a good starting point where a bit of complexity can be added to the model later.

Since many decision points are part of the same elongated street in the selected area, the agent has to traverse the entire set of nodes without much decision-making, for example, the section between decision points K, F, A, B, and C; if the agent enters this section from K, the options for it to move are either forward or backward at each successive steps till C, where it can move forward, backward and left. Here, the decision to move to a specific location is shaped mainly by the graph rather than the agent's will. To reduce this limitation while increasing the experiment's applicability, we created a synthetic environment to extend the framework for detailed analysis.

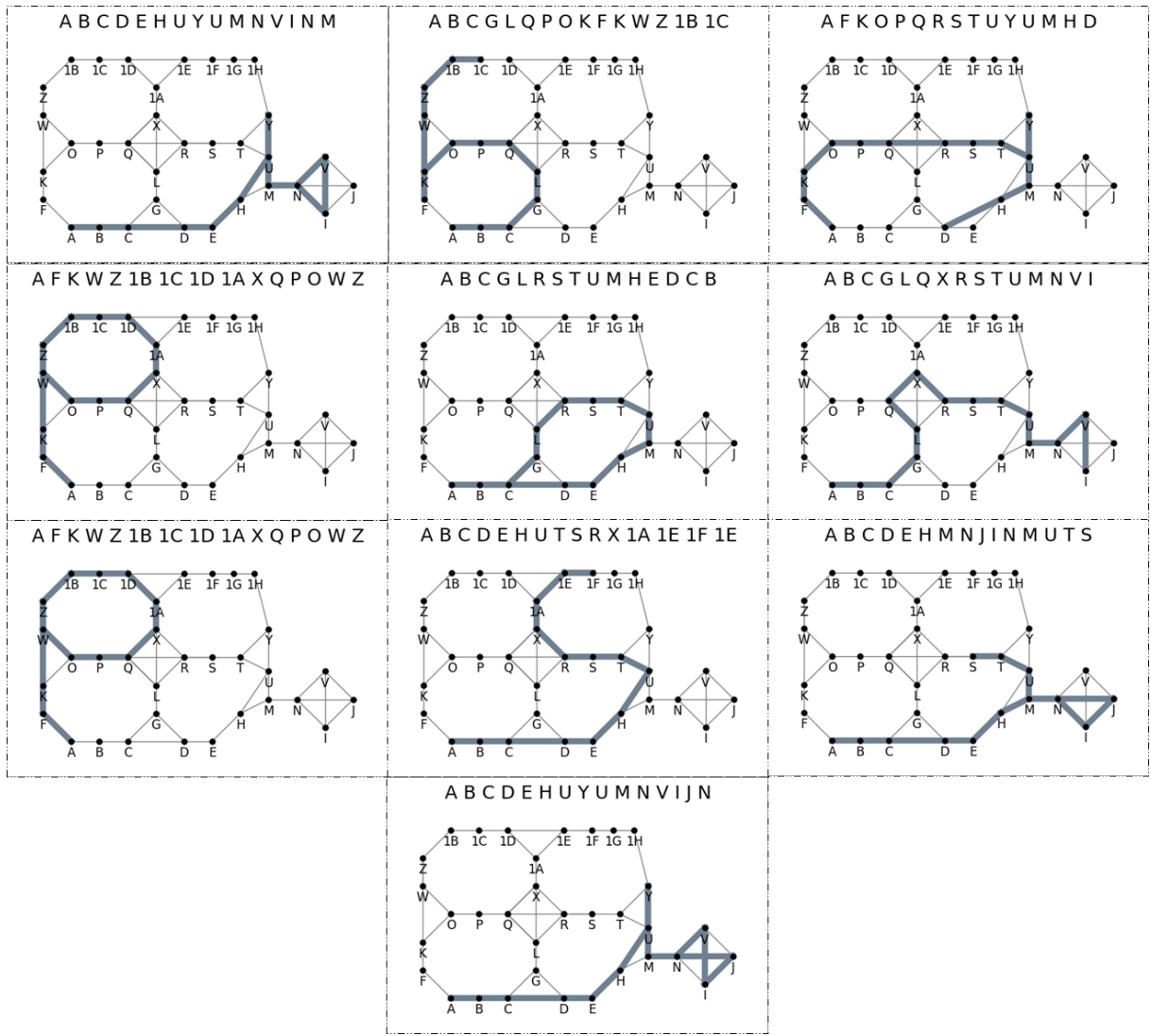

*Figure 5 Outputs from runs completed by the agent in the trial.*

## 4. Agent simulation in a synthetic environment

With the tested working set of modules and the simulation platform, we scaled the experiment to include a fixed set of decision points and introduced agent personalities and traits to see how they differ in their node selection behavior and associated strategy for decision-making. This synthetic environment includes sandbox environment creation, generation of agents' personalities, and simulations for better agent insights.

### 4.1. Synthetic environment creation

Unlike the conversion of street maps to graphs, this experiment focused on creating a synthetic set of locations or decision points. This ensures the varied multi-directional decisions that can be taken by the agent as opposed to fewer, as in the case of earlier trial attempts (Figure 6). This also shifted our focus from merely testing movement and visual modules to a more decision-making-based simulation. The environment includes a bidirectional graph of 16 nodes, each connected to its orthogonal node. The agent's starting point and the prospective final destination are marked on the map. The target location is deliberately kept at the farthest possible node to enable agents to maximize their movement. Thus, the number of minimum nodes an agent can travel before it finds its target is 5, while theoretically, there is no limit on the maximum number of nodes traversed by the agent. The rest of the framework, including memory, movement, and visual modules, is kept the same.

Similar to the graph created for the trial run, each node in the synthetic graph consists of a Google Street View URL, and the edges contain the directions. The URLs are manually selected from the broad set of URLs in Amsterdam. This step ensures enough variety in the composition and visual details of the images. From a group of 16 nodes, node A is the starting point, and the agent finishes when it successfully finds its target, "a restaurant," at node P. Nodes A, B, C, K, and L show slightly busier streets with vehicles, cycles/and or people. Nodes D, E, and H show empty, narrow roads with no people and cars, while nodes M, N, and O have a mix of parked vehicles and people. All the images have varying amounts of buildings, vegetation, and visible sky.

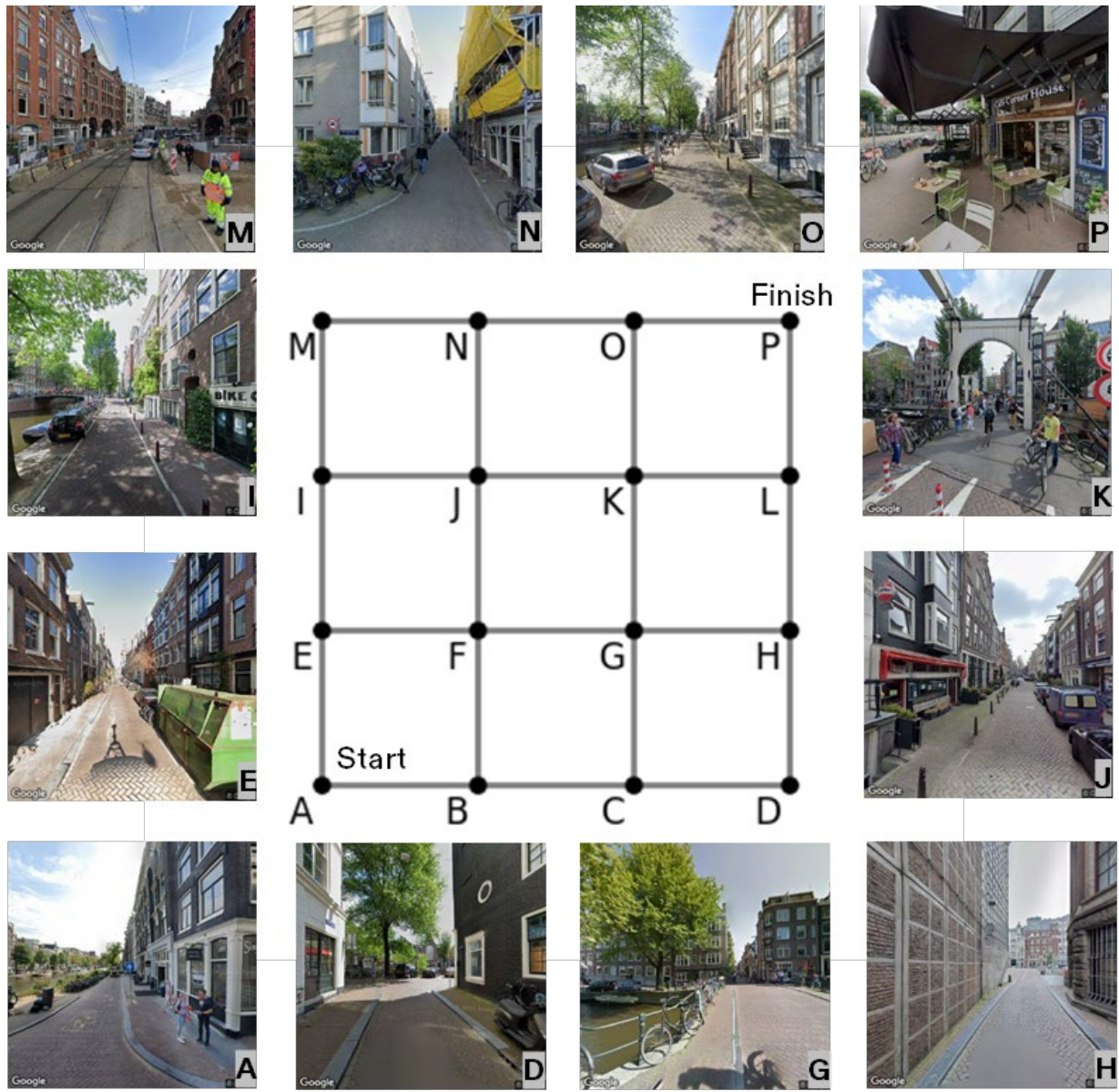

*Figure 6 Decision points and associated images in a created synthetic environment.*

## 4.2. Agent personalities

Personalities ground the agents in the simulated reality, differentiating them based on their likings, hobbies, and previous knowledge level in a particular subject. Like humans, it helps agents indulge in a behavior aligned with its backstory and prior experiences. As these Generative agents have shown to be believable proxies of human behavior, we integrate personalities into the agents to test if agents adhere to the added information regarding their decision-making in selecting directions.

We generated 10 agent personalities, each with a 150-word backstory (Figure 7), including their previous education, likes and dislikes, and passion. We also generated personality traits for each agent, which gives an immediate understanding of their qualities. These background stories and characteristics are generated with the help of ChatGPT 3.5, where the prompt specifies the creation of 10 varying personalities of different genders and age groups above 18. The agents with the personality information attached are programmed to run 10 times each. Each agent simulation is run afresh with no prior memory of the run to test the consistency of its decision in route selection.

> **Prompt:** I am creating a backstory of 10 virtual characters. These characters can be male or female over 18 years of age. The backstory should include "Personality Traits," which provide their inherent characteristics. For example: "anxious, like design, talkative." The backstory should then introduce the character and describe their hobbies and previous education in a maximum of 150 words.

**1. Emily (28)**

***Personality Traits:*** *Curious, observant, articulate, adventurous, open-minded, empathetic, detail-oriented.*
Emily is a 28-year-old with a deep appreciation for language and culture. Armed with a master's degree in Linguistics, she's a meticulous problem solver who thrives on decoding the complexities of human communication. In her free time, Emily embarks on globetrotting adventures, capturing the beauty of diverse cultures through her lens. She's an avid photographer who documents her journeys, sharing her insights through a travel blog that she curates with precision. Emily's writings reflect not only her linguistic expertise but also her passion for exploring the intricacies of the world. Her blog has garnered a dedicated following, and she takes pride in fostering connections with people from various corners of the globe.

**2. Marcus (23)**

***Personality Traits:*** *Tech-savvy, innovative, analytical, competitive, collaborative, enthusiastic, adaptable.*
Marcus, at the age of 23, is a tech-savvy person. He holds a bachelor's degree in Artificial Intelligence, which has allowed him to dive deep into the realms of programming and machine learning. Marcus is an enthusiastic video gamer who often finds himself engrossed in virtual worlds and solving complex challenges in the digital realm. His DIY electronics projects have earned him a reputation as the go-to guy for tech advice among his friends. He's a regular participant in hackathons, where he showcases his coding skills and collaborates with fellow tech enthusiasts to create innovative solutions. Marcus is the embodiment of the digital age, driven by an insatiable curiosity for emerging technologies.

**3. Sophia (27)**

***Personality Traits:*** *Empathetic, compassionate, mindful, introspective, nurturing, altruistic, patient.*
Sophia, aged 27, brings a wealth of empathy and insight to her work. With a degree in Psychology, she's adept at understanding human emotions and behavior. In her free time, Sophia practices mindfulness and yoga, seeking personal growth and inner balance. She's an advocate for mental health awareness and often volunteers at a local animal shelter, finding solace in the companionship of animals and the therapeutic benefits it brings. Sophia's generative creations often reflect her keen understanding of the human psyche, as well as her compassion for the well-being of all living beings.

**4. James (52)**

***Personality Traits:*** *Inquisitive, patient, detail-oriented, passionate, nostalgic, appreciative of tradition, adventurous.*
James, a 52-year-old, has a deep-rooted fascination with the past. Armed with a degree in History, he's a patient and detail-oriented researcher, constantly on a quest to unearth forgotten stories. When he's not immersed in historical texts, James enjoys collecting vintage books, which have become a testament to the rich tapestry of human experiences. His weekends are often spent exploring historical sites, where he immerses himself in the ambience of bygone eras, and he's an enthusiastic participant in historical reenactments. James' generative work is a testament to his passion for preserving and sharing the narratives that shape our world.

**5. Aisha (24)**

***Personality Traits:*** *Creative, expressive, passionate, melodic, versatile, confident, performance-driven.*
Aisha, at 24, has a deep passion for music. Holding a degree in Music Composition, she is a multi-talented musician, proficient in playing multiple instruments. Her true passion lies in composing her own music, and she finds immense joy in experimenting with melodies and harmonies to create compositions that reflect her innermost emotions. Aisha is a regular performer at local open mic nights, where she shares her musical creations with an appreciative audience. Her generative work is often a harmonious blend of creativity and rhythm, a reflection of her profound connection with the world of sound and melody.

**6. Carlos (35)**

***Personality Traits:*** *Energetic, competitive, analytical, sociable, disciplined, enthusiastic, team-oriented.*
Carlos, at 35, has a lifelong passion for sports. With a bachelor's degree in Sports Science, he's gained a deep understanding of the human body's physical capabilities. Carlos is an avid basketball player, relishing the thrill of competition and camaraderie on the court. When he's not shooting hoops, he can be found attending sporting events, closely analyzing athletes' performance statistics, and discussing the intricacies of the game with fellow enthusiasts. Carlos is the embodiment of the sports fan who thrives on the science of the game, analyzing and celebrating the achievements of athletes.

**7. Priya (29)**

***Personality Traits:*** *Environmentalist, outdoorsy, caring, determined, conservation-minded, resilient, community-oriented.*
Priya, aged 29, is a person with an unwavering commitment to environmental conservation. Armed with a degree in Environmental Science, she passionately dedicates her weekends to outdoor pursuits that champion the environment. Priya finds solace in hiking, trekking through the wilderness, and tending to her lush garden, where she nurtures a variety of plants. Her passion for conservation extends to volunteering at local environmental projects, where she works tirelessly to protect and preserve the natural world. Priya's generative work often echoes her love for nature, advocating for sustainability and environmental awareness.

**8. Liam (22)**

***Personality Traits:*** *Creative, gastronomic, hospitable, enthusiastic, detail-oriented, adventurous, sociable.*
Liam, a 22-year-old, has a palate that dances with flavors. Holding a degree in Culinary Arts, he's a culinary connoisseur who thrives in the kitchen. When he's not experimenting with new recipes, Liam enjoys writing about food, capturing the essence of gastronomy in his blogs. He's known for hosting delightful dinner parties for friends and family, where he showcases his culinary expertise. Liam's generative creations are a culinary celebration, often inspired by his deep love for food and his desire to share the joys of gastronomy with the world.

**9. Isabella (25)**

***Personality Traits:*** *Imaginative, literary, thoughtful, articulate, introspective, communicative, open-minded.*
Isabella, at 25, is a person with a profound love for literature. Her degree in Literature has deepened her understanding of storytelling and the power of words. Isabella is an avid reader, devouring classic novels and poetry with fervor, and she often hosts literary discussions with fellow book club enthusiasts. Her creative pursuits include writing her own stories and poetry, each a reflection of her own experiences and the emotions that flow through her. Isabella's generative work is a testament to the art of storytelling, drawing inspiration from the timeless classics that have touched her heart.

**10. Diego (30)**

***Personality Traits:*** *Analytical, strategic, mentorship-oriented, knowledgeable, disciplined, data-driven, communicative.*
Diego, aged 30, is a person with a sharp analytical mind and a degree in Economics. He thrives in the world of financial data, where he constantly analyzes markets, identifies trends, and makes informed investment decisions. When he's not immersed in the world of economics, Diego mentors and educates aspiring young investors, sharing his knowledge and helping them navigate the complex landscape of finance. Diego's generative work often carries the weight of data-driven decisions and an appreciation for the intricacies of economic systems, reflecting his dedication to the world of finance.

*Figure 7 AI-generated agent personalities generated using ChatGPT 3.5.*

*4.3. Movement*

Figure 8 shows the results from 10 cumulative runs by each Generative agent, where the width of the edge suggests the frequency of the edges used by the agent to move to their target, in other words, paths being chosen most of the time. Unlike the agent's trial version, each agent is tasked to find the restaurant present at node P and is not bounded by the number of steps taken to reach the target. Theoretically, the agent can oscillate indefinitely between a few cyclic nodes and never reach the target. However, since the paths selected, the collected visual information, and the decisions are captured by the agent's memory, agents make sufficient movement decisions to avoid these scenarios.

Figure 8 shows cumulative paths taken by each agent in 10 runs. The routes the agent considers the most are mentioned, along with the frequency of the selected path. For example, from the 10 runs, agent 1 took the path 'ABCGKLP' 3 times. The figure shows that the least traveled nodes are D, M, N, I, and E, while the most traveled include B, C, F, G, K, O, and L. These less-traveled nodes exhibit similarities in their content, such as the presence of alleys, fewer horizons, and fewer or no people and vehicles. On the contrary, the alternatives to these nodes present a more dynamic environment that most agents want to pursue to find the target. Although not precisely consistent with their choices, agents do change their decisions in different runs. However, they remain primarily stable with their choices of preferred paths throughout the runs, shown by the frequency of choosing the same route more than 3 times in 10 runs.

It can be seen that agents avoid a few nodes altogether or rarely visit if other options/directions look more probable to find the target "restaurant." Also, the paths depend on the initial decisions made by the agent to move, which can make finding and reaching these nodes difficult. Selecting one node versus the other mainly depends upon the 'target.' For instance, agents might choose different nodes if targets are 'shed,' 'ruin,' 'slum,' or 'park' based on the probability of finding these targets at subsequent nodes.

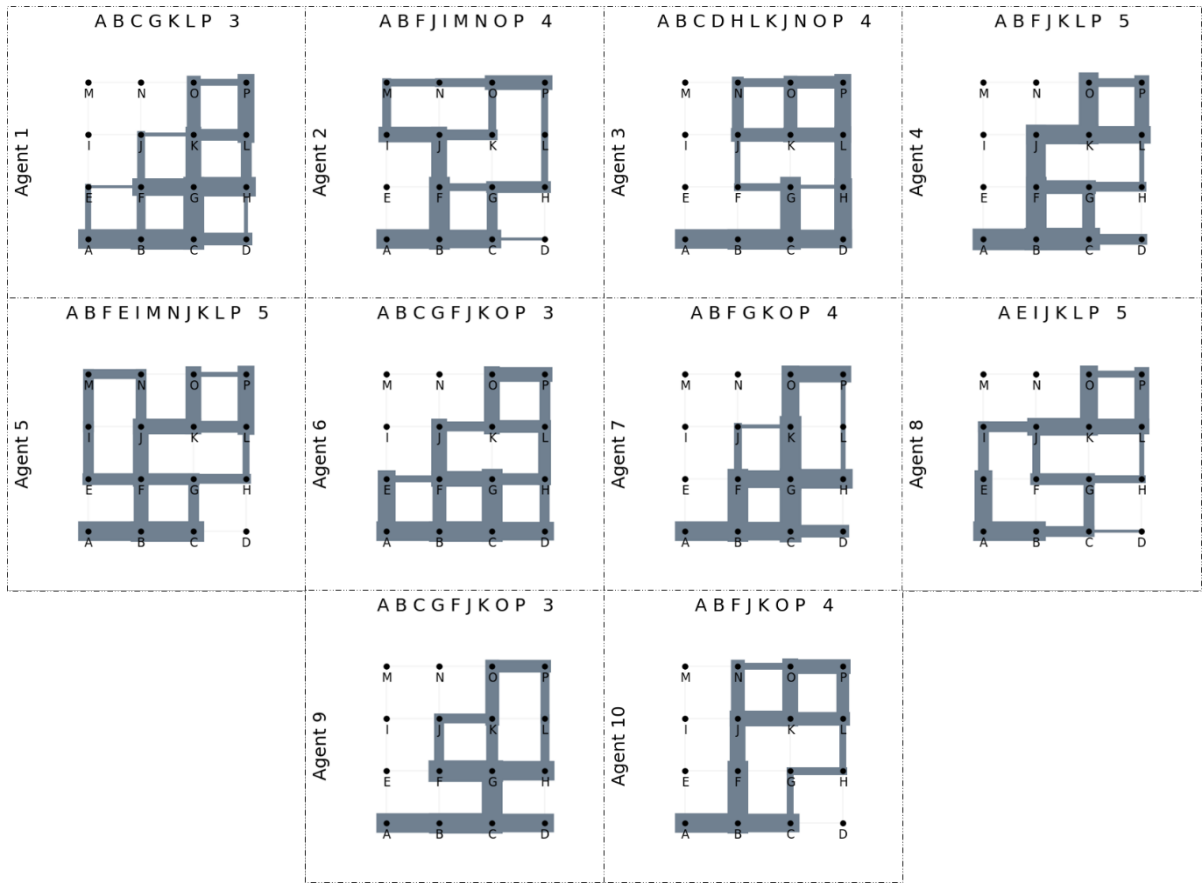

*Figure 8 Cumulative paths taken by every agent in the synthetic environment.*

While observing the results, it isn't easy to judge if the directions pursued by agents are directly based on their personalities. However, a few agents show decisions that are well-aligned to their personalities compared to the rest. Agents 1 and 3 like human conversation and chose almost all the nodes that portray people and vehicles. In contrast, Agent 7 decided on all the nodes with substantial greenery, following its backstory and personality traits of being human-like, human-conversational, and environment-loving. Such deduction on the route taken is not directly visible for all the agents; however, as they provide outputs regarding their choices, the reasons for these micro-decisions can be determined. An extensive array of choices between individual nodes is generated from 10 runs for each agent, where agents make decisions at every node and provide responses. Studying all such choices is a time-consuming task.

However, we examined one such decision between nodes B and E (Suppl 3, Suppl 4) during each agent's first simulation run. Node B comprises vegetation with a lively atmosphere with multiple people around. In contrast, Node E has less foliage and an empty street with more artificial features. Most agents preferred to go towards node B as they feel it's excellent for cultural exploration, pedestrian-friendly environment, slow and unhurried exploration, garden-friendly and community-oriented setup, and environment conducive to contemplative and imaginative thinking. Node E (Suppl 4) draws less appeal to agents' adventurous and open-minded spirit, devoid of a quieter and culturally rich atmosphere, which conflicts with the agents' desire for a tranquil and eco-conscious journey.

## 5. Perceptual responses from agents

The importance of research related to gathering inputs using visual stimuli has been fundamental in understanding human perception and behavior. These research studies experimented with visual stimuli such as photographs, drawings, and videos in various settings and a wide array of participants, including inter-cultural mix and gender and age-balanced groups. The methods used in environment/urban perception studies rapidly changed due to the sudden availability of visual datasets and AI algorithms to extract relevant information quickly. Studies such as Place-pulse (Salesses et al., 2013), Urban Gems (Quercia et al., 2014), and Streetseen (Evans-Cowley & Akar, 2015) laid the foundation for large-scale perception models that can be deployed and scaled faster than traditional ones. These studies have utilized a large group to rate/choose the images based on how safe, lively, beautiful, and boring they look (A. Dubey et al., 2016; W. Min et al., 2020; Verma et al., 2020).

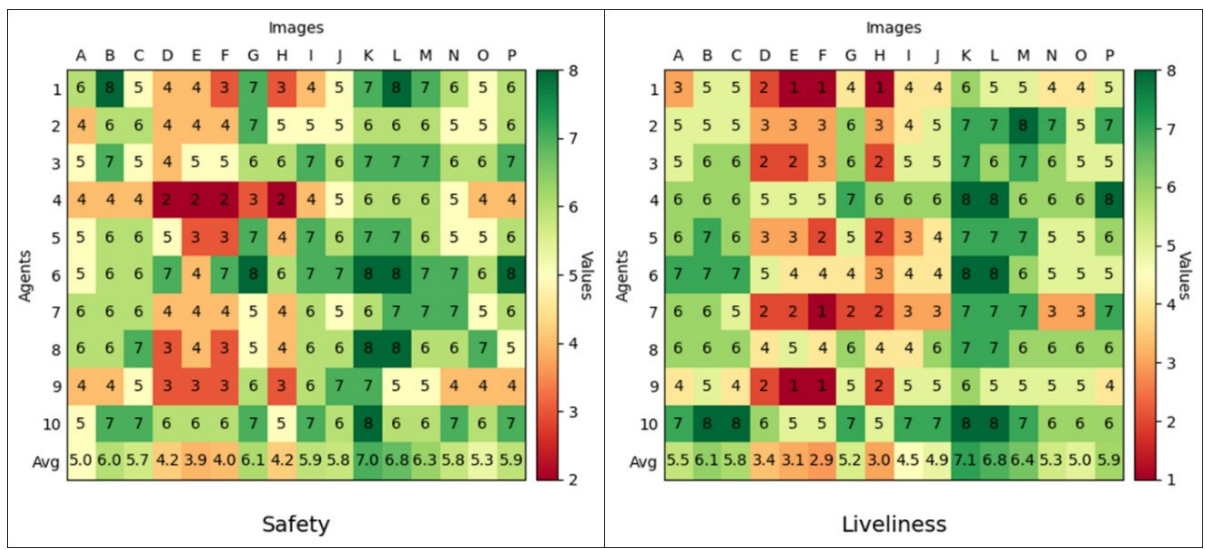

*Figure 9 Generative agents' perception of safety and liveliness of each location.*

In this experiment, we test the capability of Generative agents in replicating such participant-based trials. The experiment employs each of the 10 agents to rate all the 16 scenes used in the synthetic environment. The agent is asked to rate the images on a scale of 1 to 10 on an increasing scale of safety and liveliness. An agent's rating also details the thought process of selecting such a rating. Moreover, the agent also provides details that could be added or subtracted to make the scene a perfect 10 or 1 respectively. This experiment follows the last run of each agent; the memory, including past decisions and tasks, is still intact. Each scene from node A to P is subsequently served, and the output as a rating is shown in Figure 9.

The ratings for safety and liveliness show good segregation of the low- and high-rated images. However, the agents utilize different ranges of scale to provide ratings. None of the agents gave a rating from a minimum of 1 to a maximum of 10. Some agents dwell in the middle figures and range from 3-7, which is especially common in the safety ranking. The ranking of liveliness, however, is slightly more diverse. While agents provide micro-details regarding their choice of ranking both the attributes, a simple overview of the selections shows that most of the agents rate the alleys, scenes with no/fewer people, and scenes with overall fewer features such as nodes such as D, E, F, H and O lesser than the scenes with more variety and transient qualities such as people and vehicles in G, K, L, and M. Nodes A, B, C, I and J show somewhat mixed and intermediate ratings between these two extremes.

We closely examined a few of the outputs provided by the agents as reasons for safety and liveliness ratings (Suppl 5, Suppl 6). We selected scenes K and H for examination since these scenes are rated most safe and least safe, respectively. According to agents 1 and 6, scene K presents with openness and visibility, and the presence of individuals engaging in activities contributes positively to safety perception (Suppl 5). While scene H feels less safe due to its low visibility of sky and vegetation, the absence of people and vehicles also adds to the perception of emptiness. The agent's reasons for increasing the safety rating to 10 and reducing it to 1 are relatively similar. Almost all the agents suggest adding more lighting and community-engaging features for a perfect 10 rating while adding obstacles to hinder visibility will give a scene least rating of 1.

Similarly, we selected scenes L and F to study the reasons for liveliness given by agents (Suppl 6). The presence of people, bicycles, cars, and other elements in the scene, including the promenade and phone booth, suggested a dynamic and bustling environment in location L. Scene F is rated least lively due to the absence of people and human activity. Some suggestions to improve the liveliness are outdoor events, street vendors, vibrant colors, and organized events. These reasons are relatively consistent across all the agents; however, they differ slightly and include the agent's personality-based activity, for example, organized events, mentorship programs, and data-driven planning to enhance liveliness is consistent with agent 10's personality traits.

Agents provide varied responses regarding their decisions to choose nodes, providing ratings and the reasons to justify them. The results show the approach's stability and allow new modules and functions to be added to the framework to enhance the agent's ability to process details. However, due to the smaller environments considered in the study, researchers (us) can manually judge the results. Larger environmental boundaries may require a different strategy to simulate Generative agents and methods to analyze outputs.

## 6. Environment boundaries and modeling errors

While the simulation framework works as expected, the details, reasons, and thoughts gathered by the agents show some incorrect and invalid results. Plenty of these have come from the LLM model used and the design of the simulation framework.

The setup to execute the experiment vastly depends upon the size of the environment. The study focused on studying the application of the designed modules and analyzing the decisions by the agent, which prioritized workability versus size. However, it will be equally interesting to see how the agents perform in a larger setting, especially when the memory stream and the visual details must be engineered to deal with context windows and running costs as limiting bottlenecks. In the current setup, the context window includes the text regarding the agent's name, text describing personality and tasks, and a list of observations sorted according to recency. The observations are added at every decision node, which provides information regarding the scene at that node. Plus, it adds the data from the images from all the related directions, aiding agents' decision-making. The larger environment would mean a growing list of observations, agents' responses, and decisions, which might overshoot the capacity of the backbone LLM being used for the agent. The LLMs with more oversized context windows are available, but the model performance diminishes as the window length increases.

While the simulation works as intended, with all the modules working as programmed, it can be seen that the agents are sometimes driven by inherent irregularities or presumptions. This behavior is occasionally erratic and non-consistent, and the inherent bias in preferences can go beyond what agent personalities offer. Since we involve LLM to undertake cognitive tasks such as movement, for which it is not trained and modeled, the instances where the movement shows abnormal human behavior are plenty, the prime examples being crossing the same node twice in quick succession or retreating from the initially chosen path or circling around the same couple of nodes. This happens even when the agent has stored the information regarding already visited nodes and their details in the context window, and it is also visible when the same agent does a separate run.

While the memory storage function adopted directly from the study (Park et al., 2023) works well with the experiments, the rating function, which rates the observations on a scale of 1 to 10 based on importance, puts a lot of emphasis on the agent's assumed knowledge of the two extremes. It returns values that might be nonsensical, depend on random chances, or irrelevant to the topic. The memory might work well in environments with apparent extremes like those used in the study, such as when a scene has water, a fountain, or a starkly dimly lit alley, but might struggle in homogeneous environments such as regular residential neighborhoods having similar visual qualities. Similarly, the model assigns the agent to rate images for safety and liveliness. The agents provide their understanding and the reason for choosing a specific value. However, the current tools cannot evaluate whether that value is justified for the received reason.

Movement incurs the passage of time, which is essential to consider to restrict agents from exploring the surroundings endlessly. Further, since the synthetic environment and the small study area have limited visual information, the function of time vs distance is not studied. The physical cost of moving akin to humans is challenging to emulate in virtual models. Similarly, we humans do not judge the subsequent scenes individually but as a continuous stream of visual information where features we identify get added to our memory over the existing visual information common in these scenes (Lindsay & Norman, 2013). We tried replicating the same effect in a virtual medium by creating a sub-memory base for all the visual details calculated by the visual agent, where it reasons within itself when new image information is encountered to see if the new information consists of novel details. The module then adds up only the novel information captured from the visual realm over the previous information from a series of images up to this point. The module worked as expected; however, the agent could not effectively decide on the relevant information or the details to use while storing the information on the new experienced scene. This exposes the capability of the LLM agents to process minute details and calculate their importance in the overall scene.

In the current setup, the experiments utilize only a primary visual input from DL models, which provide an overview of the scene rather than detailed nuances of each identified feature. We initially used a more extensive collection of visual DL models to extract visual elements in detail; however, it generated a significant amount of knowledge for the agent, preventing it from making quick and informed decisions. While a general-purpose simulation is still workable, the details are necessary to guide various choices, such as (a) whether to go in a particular direction if it has smooth pavement instead of cobblestones or textured, (b) street furniture to suggest more pedestrians and recreation zones, (c) The idea of street widths to determine if it is a main street, traffic signals, and signs that restrict access or movement, (d) Bus and tram stations, which induce the idea of the availability of other transportation modes, and (e) The types of buildings and urban spaces suggest the main market area, commercial hub, or residential neighborhood.

## 7. Discussion

This study generated a quick testbed to analyze the potential functions possible with the Generative agent, which can aid urban perception studies. The flexibility of the agent to include a list of modules enhanced the scope of this study, without which it would have been limited to generating text-based Q&A and general information. Throughout the study, it is seen that these Generative agents can understand regular conversations, create thoughts, and provide solutions given their scope. Before such a mechanism, the tools had to be given proper hard-coded instructions to obtain meaningful outputs. A significant addition of capability is the role-playing behavior of the agent, where it can generate different personalities and act according to the persona while executing various tasks. Adding human-like qualities allows more exciting research involving human behavior and cognition. Given the scale and complexities of living spaces in cities, these agents may pave paths for virtual surveys, where they can effectively rate the surroundings, simultaneously accounting for, e.g., demographics and gender roles as they assume different personalities.

This study utilizes visual foundation models to extract details from street-view images. Since the streets are constantly changing, the real-time assessment of urban areas is a prime challenge. The multimodality and temporality (Spence, 2020; Verma et al., 2019) affect the perceived nuances of the environment; however, their assessment has been an ever-challenging phenomenon due to a lack of datasets and human-based decision-making, such as Generative agents. Further, since an agent cannot inherently see and hear, multiple models assist it in understanding the visual and auditory realm. However, the clarity and content of these descriptions affect the agent's understanding of the place. For instance, the fundamental visual models are trained to output details as values, such as the number of vehicles or pedestrians in the street; however, humans do not evaluate scenes with values. Humans process visual information by registering novel feature instances while leaving out common elements in successive scenes. The information registration also considers auditory, olfactory, and tactile senses, uniquely amalgamating observed space. Recreating a similar experience virtually is still impossible with the current tools at hand, and emulation of these senses by providing a caption or explanation of these complex scenes might overburden the agent. The studies (Driess et al., 2023) have started using robotic devices that observe their environment in real-time to react and process information; however, these are mainly in controlled spaces or have specific goals. A real-time Generative agent aimed to collect perceptual thoughts and decisions based on actual urban scenarios might be the best survey participants in the future.

Compared to other tools, a critical distinction of Generative agents is their inherent capability to generate thoughts and the mechanism that aligns itself with solving the presented problem. However, these qualities also pose difficulties in understanding the agents' decision of whether they are taken by evaluating observations or random chance. Retrieval of stored memory plays a part in decision-making in agents; however, the agents may or may not adhere to it. Finding the root causes of their decisions becomes challenging due to multiple interconnections between their personality traits and recent observations that are difficult to study in isolation.

A significant limitation to the Generative agent-driven research is the size of LLMs, the service costs, and the internet dependence. The prices are calculated per token, roughly a small group of characters or alphabets. Since the agent includes a lot of prompt text to ensure the supply of correct context for the agent to work effectively, the tokens supplied to the APIs are enormous which incur significant costs (Kaddour et al., 2023). While open-source LLMs that offer performance levels similar to those used in the study have been released, they still must be hosted on servers with more specifications than consumer-grade workstations. However, models capable of running on low computational resources have been introduced but suffer from inaccuracies. Finetuning with context-relevant datasets with such models might be a temporary fix (Shanahan, 2022).

## 8. Conclusion

The study focused on utilizing large language models (LLMs) as Generative agents, then deployed virtually to emulate human perception and feedback through virtual movement and visual modules. While LLMs are being utilized in multiple tasks across a broad set of domains, this study has focused on (1) testing the capability of the present state-of-the-art models in cognitive tasks of virtual movement and its corresponding decisions and (2) identifying the gaps in the current mechanisms that are related to LLMs, the model architecture of Generative agents, the functionality of the platform and degree of mismatch between real and virtual world.

The study created experiments involving the usage of agents with different personalities and their capability to navigate urban neighborhoods with the help of real-world scenes. The movement module included bidirectional graphs, where nodes acted as decision points with directional information stored on the edges. The agents are given a set of ten different personalities to see if their decision-making regarding the chosen path changes according to their past experiences, likes, and dislikes. The decisions to move depend on the inputs each agent gathers from the surroundings, which affects the decision-making and comprehension of the world around them. Further, these agents are also employed to rate the scenes according to their perception of safety and liveliness.

Although the experiments provided insights into the capabilities of LLMs in general, it has also been clear that the current state of these models performs well only when providing enough guidance through prompts and structure. Further, the possibility of attaching detailed observations and knowledge that the agent utilizes for decision-making increases its information load and hampers its functionality. While the research on using multimodality with LLM is ongoing, we expect this domain will continue to grow, and fundamental changes in the model architecture and training mechanisms will include datasets and processes to have even non-linguistic knowledge such as cognition, planning, and embodiment.

Studying behavior and perceptions has vast importance in environmental psychology and perception. Adopting current state-of-the-art models to perform these tasks provides us with the latest shortcomings and capabilities of these models and helps us chart new methods to best utilize these in experiments related to environment perception and allied fields.

**Funding**
This research was funded by the Ministry of Science and Culture of Lower Saxony and by the Volkswagen Foundation under Grant 94957.

**Data Availability Statement**
All the datasets used in the study are cited in the text.

**Supplementary 1: The working of the created simulation framework.**

**> Simulation start**
**> Creation of a Generative agent**
> Name: Max (age: 25)
Innate traits: N/A
Backstory and personality: N/A
Max is intelligent, hardworking, and reliable.
It is November 02, 2023, 12:43:21 PM.
Max's status: Want to explore the city and find a tree-house around to write a story about it.

**> Initial observation of the agent using the Visual inference module**
> First observation ==> Max comes out of his home at location A. The directions in which Max can move are forward, right. In the forward direction he sees The scene is densely filled with buildings and vegetation. It seems like a residential neighborhood, a street, a canal/urban area, an alley, and a courtyard. It has a man-made setting illuminated by natural light. It's an open area, sunny but with no horizon visible, and it has a lot of foliage and trees. There are four cars and a bicycle in the vicinity. In the right direction The environment is dominated by buildings and roads, with a low presence of sky and vegetation. The location can be characterized as a street, possibly an alley and has features similar to a medina, outdoor parking garage, and an arcade. The setting is man-made, under natural light, and suitable for driving and biking. It's an open area, but there's no horizon. There's a potted plant in the vicinity. Asphalt, pavement, and glass are also present.

**> Prompt given to the agent to provide its reaction on which node it needs to move to**
> Max should react to the observation, and if so, what should be the next step for Max to move from the given directions? Respond by only one direction Max should focus on moving to. The potential responses should have one of the four directions: forward, backward, left, and right. However, Max should see which directions are available for him and most likely to have a tree-house based on the observation and respond accordingly. Also the Max should justify why it wishes to go in that particular direction and also why not the other direction based on his/her personality.
REACT: Max's reaction in the format: Max wish to go to .... direction because ...., Max does not wish to go to .... direction because . . .. Be precise and write only the things that you know from the context. Do not mindlessly select any direction, reason before you choose.

**> Agent provides the reaction, including the reason for choosing the specific node**
> Max wish to go to the forward direction because based on the observation, it seems to be a street which is most likely to have tree-house, given great amount of foliage and trees and setting being a residential meighborhood illuminated by a natural light. The presence of cars and bicycle shows a lively environment. Tommie does not wish to go to the right direction because although there are multiple possibilities mentioned, the scene is man made with large features of streets and alley which seems to not have tree-house located there.

**> Agent observes the new location to find the target by evaluating additional GSV imageries**
> Next_observation ==> After coming from location A, Max moved forward to B, he looks on right and left sides of the street to find the tree-house. The scenes appears to be a combination of a bank vault, an outdoor doorway, a barn door, an elevator door, and a shopfront, combination of a campus, a hospital, an embassy, a residential neighborhood, and a canal/urban area. Overall the scene is densely filled with buildings and vegetation. It seems like a residential neighborhood, a street, a canal/urban area, an alley, and a courtyard. It has a man-made setting illuminated by natural light. It's an open area, sunny but with no horizon visible, and it has a lot of foliage and trees. There are four cars and a bicycle in the vicinity. No sign of tree-house yet.

**> Prompt given to the agent to rate the observation above on the scale of importance**
> On the scale of 1 to 10, where 1 is purely regular street details which one can see in regular city streets (e.g., mundane urban setting characterized by a sparse population and minimal vehicular activity. Streets are devoid of vibrant elements, with limited greenery and visible sky. Lack of noteworthy establishments like shops, cafeterias and landmarks) and 10 is extremely poignant (e.g., captivating urban scene which features a bustling environment with a diverse crowd and active vehicular presence. The scene composition includes a harmonious blend of greenery, buildings, and a dynamic sky. Streets are animated with life, showcasing variety of establishments. The overall composition which creates a memorable experience.), rate the likely poignancy of the following piece of memory.

**> Saved memory**
>- [November 02, 2023, 12:44:15 PM] After leaving location A, Max proceeded forward to B, thoroughly examining both the right and left sides of the street in search of a tree-house. The scene at B comprised various elements, including a bank vault, outdoor doorway, barn door, elevator door, and shopfront. The environment was a blend of a campus, hospital, embassy, residential neighborhood, and canal/urban area. The overall setting was characterized by dense buildings and vegetation, resembling a residential neighborhood, street, canal/urban area, alley, and courtyard. It exhibited a man-made environment with natural light, an open area, a sunny atmosphere, and abundant foliage and trees. Despite the thorough exploration, Max did not spot a tree-house in this location. Notable details included the presence of four cars and a bicycle in the vicinity, contributing to the dynamic urban landscape observed by Max.

*Suppl 1 The working of the created simulation framework.*

> **Second observation of the agent using Visual inference module**
> Now he has to decide on moving further. He can only go to these directions: backward, forward. He came here at B from backward direction. So he already saw what is there in that direction.
On the forward direction, The scene is dominated by buildings and roads, with vegetation and sky playing less prominent roles. It could be a residential neighborhood, an outdoor parking garage, a loading dock, or an industrial area, but it has street-like qualities as well. The scene is man-made and lit by natural light. The area is open, but there is no visible horizon. There are several elements suggesting transport, such as glass, pavement, and five cars and three bicycles. It's a sunny day.

> **Prompt given to the agent to provide its reaction on which node it needs to move to**
> Max should react to the observation, and if so, what should be the next step for Max to move from the given directions? Respond by only one direction Max should focus on moving to. The potential responses should have one of the four directions: forward, backward, left, and right. However, Max should see which directions are available for him and most likely to have a tree-house based on the observation and respond accordingly. Also the Max should justify why it wishes to go in that particular direction and also why not the other direction based on his/her personality.
REACT: Max's reaction in the format: Max wish to go to .... direction because ...., Max does not wish to go to .... direction because . . .. Be precise and write only the things that you know from the context. Do not mindlessly select any direction, reason before you choose.

> **Agent provides the reaction, including the reason for choosing the specific node**
> Max wishes to go in the forward direction because based on the scene analysis, it seems that moving forward will lead him towards the street in a residential neighborhood. This direction is more likely to have tree-house and other establishments for Max to explore. While the scene has less vegetation and sky, Max is still positive of finding tree house there. Max does not wish to go in the backward direction because he already came from there and has seen what is in that direction.

> **Agent observes the new location to find the target by evaluating additional GSV imageries**
> Next_observation ==> After coming from location B, Max moved forward to C, he looks on right and left sides of the street to find the tree-house. The scenes can be classified as combination of a fire escape, a jail cell, a barndoor, a shopfront, and an outdoor doorway, combination of an embassy, a parking lot, a canal/urban area, a junkyard, and a hospital, but no tree-house. Overall, The scene is dominated by buildings and roads, with vegetation and sky playing less prominent roles. It could be a residential neighborhood, an outdoor parking garage, a loading dock, or an industrial area, but it has street-like qualities as well. The scene is man-made and lit by natural light. The area is open, but there is no visible horizon. There are several elements suggesting transport, such as glass, pavement, and seven cars and three bicycles. It's a sunny day.

> **Prompt given to the agent to rate the observation above on the scale of importance**
> On the scale of 1 to 10, where 1 is purely regular street details which one can see in regular city streets (e.g., mundane urban setting characterized by a sparse population and minimal vehicular activity. Streets are devoid of vibrant elements, with limited greenery and visible sky. Lack of noteworthy establishments like shops, cafeterias and landmarks) and 10 is extremely poignant (e.g., captivating urban scene which features a bustling environment with a diverse crowd and active vehicular presence. The scene composition includes a harmonious blend of greenery, buildings, and a dynamic sky. Streets are animated with life, showcasing variety of establishments. The overall composition which creates a memorable experience.), rate the likely poignancy of the following piece of memory.

> **Saved memory**
>- [November 02, 2023, 12:44:15 PM] […]
 - [November 02, 2023, 12:45:01 PM] Following Max's departure from location B, he proceeded forward to C, actively surveying both the right and left sides of the street in hopes of spotting a tree-house. The scenes at C encompassed a variety of elements, including a fire escape, jail cell, barndoor, shopfront, and outdoor doorway, presenting a combination of features from an embassy, parking lot, canal/urban area, junkyard, and hospital. Despite the diverse array of components, no tree-house was observed. The overall scene was characterized by a dominance of buildings and roads, with vegetation and sky playing lesser roles. It exhibited qualities reminiscent of a residential neighborhood, outdoor parking garage, loading dock, or industrial area, with street-like features. The environment was man-made, lit by natural light, and open, though lacking a visible horizon. Transport elements such as glass, pavement, five cars, and three bicycles were present, contributing to the scene's sunny and transport-oriented ambiance.

**Supplementary 2: Results from interviewing the agent after a complete simulation run during testing the framework. The agent followed the path: A-B-C-D-E-H-U-Y-U-M-N-V-I-N-M.**

**Q. What directions did you take from your starting point A to all other subsequent locations?**
Starting from location A, I observed and moved to various destinations with specific directions. From A to B, I moved forward. The decision was then made to proceed forward due to the potential for finding a tree-house in the busy urban area. Moving from B to C, I chose to go forward again, exploring a serene residential area near a canal, with lush vegetation and no specific object detections. In the subsequent journeys from C to D, M to N, D to E, E to H, and H to M, I consistently opted to go forward, guided by observations of diverse urban environments, streets, promenades, and canal/urban areas. This direction was favored due to the likelihood of finding a tree-house in areas exhibiting a mix of residential and commercial characteristics. Throughout the exploration, I strategically avoided revisiting previously explored directions, ensuring efficient progress towards the goal.

**Q. If you wish to come back, How will you come back from point M back to A? Describe step-by-step in detail.**
To return from point M back to point A, I would need to retrace my steps in the reverse order. Starting from M, the last observed location, I would need to move backward through each location visited sequentially. Following the provided information, the reverse order would be N to M, then H to E, D to C, and finally C to B. From B, I would move backward to A, completing the journey in the opposite direction. It's essential to follow the reverse path accurately to revisit each location and ensure a safe and efficient return to the starting point.

**Q. Did you find the tree-house on the way?**
The provided information about my journey does not explicitly mention whether I found the tree-house during the described observations and movements. The focus has been on my decision-making process based on scene observations and recommendations for the next direction to take. The information emphasizes my exploration of diverse urban environments, streets, promenades, and canal/urban areas in the quest for finding a tree-house, but the actual discovery of the tree-house is not confirmed in the knowledge I have. If there are additional details about my findings or the outcome of the journey, it would be helpful to include them to determine whether I successfully found the tree-house.

**Q. Why did you choose the specific directions, when you could have taken other directions?**
The specific directions chosen by me in the described journey were influenced by the observations made at each location, my goal of finding a tree-house, and the characteristics of the scenes. I consistently opted to move forward based on the analysis of the environment, which included factors such as the presence of high buildings, vegetation, roads, sidewalks, and indications of a diverse urban environment. The decision to move forward was often guided by the belief that proceeding along streets, promenades, or canal/urban areas in diverse urban environments would increase the likelihood of finding a tree-house. I also strategically avoided revisiting previously explored directions, such as backward, left, or right, presumably to optimize the exploration process and focus on unexplored areas that held potential for the desired tree-house. In summary, my choices in direction were driven by a combination of environmental cues, the goal of finding a tree-house, and a strategic approach to avoid redundant exploration.

*Suppl 2 Results from interviewing agent after a simulation run.*

**Supplementary 3: The outputs from the visual inference module for the selected scenes in the synthetic environment.**

**B**
The scene is characterized by high presence of buildings and roads, and abundant vegetation, with few poles and fences. It is located on a street that also functions as a promenade, surrounded by a residential neighborhood, an embassy, and a campus. The area is man-made and naturally lit, featuring an open layout with no visible horizon, designed for driving and biking on asphalt and pavement, and dotted with trees. There are 12 individuals, some of whom are carrying handbags, along with 2 bicycles and a motorcycle.

**E**
The scene is characterized by a high presence of buildings and roads, with low sky and vegetation visibility. It appears to be on a busy street with potential parking, indoor bus station, residential neighborhood, and train station platform in the vicinity. The man-made structures are under natural light in an open area designed for driving and transporting on asphalt and pavement. There is a notable presence of vehicles with two cars and two trucks, along with a bicycle and a motorcycle.

**F**
The scene is mostly filled with buildings and abundant vegetation, while the sky is not much visible. It seems to be a residential neighborhood with alleys and possibly a roof garden, with occasional presence of a street. This man-made area is naturally lit and appears to be an open area, but the horizon is not visible. It's paved with shrubbery, leaves, and foliage. There is no presence of people in the scene.

**H**
The scene is characterized by high presence of buildings and roads, with low sky and vegetation visibility. It appears to be an alley with elements of a medina, loading dock, jail cell, and arcade somewhat present. The scene is man-made, with natural light and both horizontal and vertical components that give it a sense of depth. It's a mostly open area, with some enclosed portions. It's paved and suitable for biking. There are no people and vehicle in the scene.

**K**
The scene has a high presence of roads, while the sky, buildings, and vegetation are average. The presence of a fence and a wall is low. It can be identified as an aqueduct, water park, playground, promenade, or arch with some possibility. This man-made area is under natural light, suitable for touring and driving, and appears to be an open, sunny area, but there is no horizon visible. There are also trees and metal elements in the scene. There are 6 people, 5 bicycles, and 2 cars present.

**L**
The scene is predominantly occupied by buildings and roads, with low sky and vegetation visibility. It resembles a street, with elements of a promenade, phone booth, outdoor parking garage, and hotel. This man-made place is brightly lit by natural light, suitable for driving and biking on asphalt and pavement, and appears to be an open, sunny area, although there is no horizon in sight. It comprises 7 people, 2 cars, 2 bicycles, a backpack, and a handbag.

*Suppl 3 The outputs from the visual inference module for the selected scenes.*

**Supplementary 4: Agents selection preference of scene B vs E.**

**Agent 1 :** Emily wish to go to right direction **(B)** because the of its lush vegetation, promenade atmosphere, and an open layout designed for biking. The residential neighborhood, embassy, and campus surrounding the area align with her love for cultural exploration. The limited vehicle presence and focus on individuals and bicycles create a pedestrian-friendly environment, resonating with Emily's preference for relaxed and nature-rich settings. Emily does not wish to go to forward direction **(E)** because in contrast, busy street, emphasis on vehicles, and less visible greenery make it less appealing to Emily's adventurous and open-minded spirit. The vibrant and diverse ambiance of the right direction makes it her preferred choice for finding a restaurant.

**Agent 4 :** James wish to go to right direction **(B)** because of his deep-rooted fascination for history. The open layout devoid of a visible horizon appeals to his patient and detail-oriented nature, providing ample space for exploration. The historical resonance of the promenade, surrounded by a residential neighborhood, embassy, and campus, resonates with James's passion for uncovering forgotten stories. The limited vehicle presence and focus on individuals and bicycles create a pedestrian-friendly environment, aligning with James's preference for unhurried exploration. James does not wish to go to forward direction **(E)** because busy street and emphasis on vehicles may feel less conducive to James's historical pursuits, where a quieter and culturally rich atmosphere is essential for his restaurant-seeking journey.

**Agent 6 :** Carlos wish to go to forward direction **(E)** because of his lifelong passion for sports. The busy street, potential parking, and train station platform align with his energetic and competitive nature, providing a dynamic backdrop for his exploration. The emphasis on driving and transporting on asphalt and pavement resonates with Carlos's deep understanding of sports science and the physical capabilities of the human body. The notable presence of vehicles, including cars and trucks, adds to the vibrant atmosphere that Carlos thrives in. Carlos does not wish to go to right direction **(B)** as the scene focus on individuals, bicycles, and a motorcycle may feel less connected to the sports-oriented environment that Carlos seeks for his restaurant-seeking journey.

**Agent 7 :** Priya wish to go to right direction **(B)** because she is committed to environmental conservation. The lush vegetation, open layout, and limited vehicle presence align with her dedication to sustainability and nature. The outdoor-focused environment, featuring hiking possibilities and a garden-friendly setting, resonates with Priya's passion for protecting and preserving the natural world. The cultural diversity of the promenade surrounded by a residential neighborhood, embassy, and campus provides a holistic experience in line with Priya's community-oriented mindset. Priya does not wish to go to forward direction **(E)** because busy street, vehicle emphasis, and potentially hectic atmosphere may conflict with Priya's desire for a tranquil and eco-conscious restaurant-seeking journey.

**Agent 9 :** Isabella wish to go to right direction **(B)** because of her profound love for literature. The lush vegetation and open layout devoid of a visible horizon provide an environment conducive to contemplative and imaginative thinking, aligning with Isabella's literary and thoughtful nature. The residential neighborhood, embassy, and campus surrounding the promenade offer a culturally rich and diverse backdrop, perfect for someone who appreciates the nuances of storytelling. The limited vehicle presence and focus on individuals, bicycles, and a motorcycle create a serene and pedestrian-friendly atmosphere, resonating with Isabella's desire for unhurried exploration. Isabella does not wish to go to forward direction **(E)** because busy street and emphasis on vehicles may feel less conducive to Isabella's literary pursuits, where a quieter and culturally rich ambiance is essential for her restaurant-seeking journey.

*Suppl 4 Agents selection preference of scene B vs E.*

**Supplementary 5: Safety ratings for selected agents and images.**

**Safety rating for scene K**

**Agent 1 :** Emily's Safety Rating for Scene K is 7. The reason for that is Emily perceives the scene as relatively safe. The open, sunny environment and the low presence of fences and walls contribute to a sense of openness and visibility, enhancing the overall feeling of safety. The presence of individuals engaging in various activities, including cycling and driving, suggests a lively and active atmosphere, which contributes positively to the safety perception.
To elevate the safety rating to 10, Emily suggests increasing the presence of visible security measures such as well-lit paths, surveillance cameras, and more people for increased social presence.
To decrease the safety rating to 1, Emily suggests adding obstacles that obstruct visibility, reducing lighting, and increasing the presence of secluded areas.

**Agent 6 :** Carlos's Safety Rating for Scene K is 8. The reason for that is Carlos perceives the scene as moderately safe. The open and sunny environment is positive, but the low presence of visible security features and the mix of individuals, bicycles, and cars may raise some safety concerns, especially in a potentially busy setting.
To increase the safety rating to 10, Carlos suggests adding more lighting, security patrols, and clearly marked pathways. Increasing the presence of visible security measures, such as surveillance cameras, would contribute to a safer environment.
To decrease the safety rating to 1, Carlos suggests reducing lighting, removing visible security elements, and increasing the presence of obstacles that hinder visibility.

**Safety rating for scene H**

**Agent 1 :** Emily's Safety Rating for Scene K is 3. The reason for that is Emily perceives the scene as less safe due to the low visibility of the sky and vegetation, and the presence of elements like a jail cell and arcade that may contribute to a potentially less secure atmosphere. The absence of people and vehicles also adds to the perception of emptiness, potentially reducing the feeling of safety.
To increase the safety rating to 10, Emily suggests adding more lighting, visible security elements, and greenery. Incorporating community-engaging features and surveillance cameras would contribute to a safer environment.
To decrease the safety rating to 1, Emily suggests adding obstacles that hinder visibility, reducing lighting, and incorporating elements that create a sense of seclusion.

**Agent 6 :** Carlos's Safety Rating for Scene K is 6. The reason for that is Carlos perceives the scene as less safe, primarily due to the low visibility of the sky and vegetation. The absence of people and vehicles may create a sense of emptiness, but the natural light and openness mitigate some safety concerns.
To increase the safety rating to 10, Carlos suggests adding more lighting, visible security elements, and greenery. Incorporating community-engaging features and surveillance cameras would contribute to a safer environment.
To decrease the safety rating to 1, Carlos suggests adding obstacles that hinder visibility, reducing lighting, and incorporating elements that create a sense of seclusion.

*Suppl 5 Safety ratings for the selected agents and images.*

**Supplementary 6: Liveliness ratings for selected agents and images.**

**Liveliness rating for scene L**

**Agent 7** : Priya's Liveliness Rating for Scene L is 7. The reason for that is Priya perceives scene as lively. The presence of people, cars, bicycles, and various elements like a promenade and phone booth suggests a dynamic and bustling environment. The presence of people, cars, bicycles, and various elements like a promenade and phone booth align with her appreciation for dynamic outdoor spaces.
To increase the liveliness rating to 10, Priya suggests enhancing the liveliness by incorporating more community-focused elements, such as outdoor events, green spaces, and opportunities for community engagement. This aligns with her caring and conservation-minded personality, making the scene more vibrant and community-centric.
To decrease the liveliness rating to 1, Priya suggests reducing the presence of people, cars, bicycles, and elements that contribute to a busy street scene. Removing the backpack, handbag, and any potential signs of life would decrease the overall liveliness.

**Agent 10** : Diego's Liveliness Rating for Scene L is 8. The reason for that is Diego perceives Scene L as lively. The presence of people, cars, bicycles, and various elements like a promenade and phone booth align with his analytical approach to dynamic environments.
To increase the liveliness rating to 10, Diego suggests enhancing the liveliness by incorporating strategic elements, such as organized events, mentorship programs, and data-driven planning. This aligns with his disciplined and data-driven personality, making the scene more vibrant and strategically engaging.
To decrease the liveliness rating to 1, Diego suggests reducing the presence of people, cars, bicycles, and elements that contribute to a busy street scene. Removing the backpack, handbag, and any potential signs of life would decrease the overall liveliness.

**Liveliness rating for scene F**

**Agent 7** : Priya's Liveliness Rating for Scene F is 1. The reason for that is Priya perceives the scene as least lively due to the absence of people. Priya, being environmentally conscious and community-oriented, finds adding eco-friendly features like a community garden, recycling stations, or environmental art installations to enhance the liveliness while aligning with her caring and conservation-minded personality.
To increase the liveliness rating to 10, Priya suggests adding more elements like people engaging in various activities, outdoor events, and street vendors. Incorporating community-engaging features and vibrant colors would contribute to a more lively environment.
To decrease the liveliness rating to 1, Priya suggests reducing the presence of buildings, vegetation, and elements that contribute to a residential atmosphere. Removing the bicycle and any potential signs of life would decrease the overall liveliness.

**Agent 10** : Diego's Liveliness Rating for Scene F is 5. The reason for that is Diego perceives Scene F as less lively due to the absence of people. While the presence of buildings, abundant vegetation, and a bicycle suggests a dynamic environment, the lack of human activity significantly reduces the overall liveliness. He suggests incorporating strategic elements like organized events, and data-driven planning to enhance the overall liveliness, aligning with his analytical and disciplined personality.
To increase the liveliness rating to 10, Diego suggests adding more elements like people engaging in various activities, outdoor events, and street vendors. Incorporating community-engaging features and vibrant colors would contribute to a more lively environment.
To decrease the liveliness rating to 1, Diego suggests reducing the presence of buildings, vegetation, and elements that contribute to a residential atmosphere. Removing the bicycle and any potential signs of life would decrease the overall liveliness.

*Suppl 6 Liveliness ratings for selected agents and images.*